\documentstyle[12pt]{article}
\begin{document}

\begin{center}
{\Large\bf GALILEAN SYMMETRY IN A NONABELIAN CHERN SIMONS MATTER
SYSTEM}
\vskip 1in

{\bf R. Banerjee}\footnote{e-mail: narayan@bose.ernet.in}\\
{\normalsize S.N.Bose National Centre for Basic Sciences}\\
{\normalsize Block JD, Sector III, Salt Lake City,
Calcutta 700091, India}\\[.5cm]
{\normalsize and}\\[.5cm]
{\bf P. Mukherjee}\\
{\normalsize A.B.N. Seal College}\\
{\normalsize Cooch Behar, West Bengal}\\
{\normalsize India}\\
\end{center}

\begin{abstract}
We study the Galilean symmetry in a nonrelativistic model,
recently advanced by Bak, Jackiw and Pi, involving the coupling
of a nonabelian Chern-Simons term with matter fields. The
validity of the Galilean algebra on the constraint surface is
demonstrated in the gauge independent formalism. Then the reduced
space formulation is discussed in the axial gauge using the
symplectic method. An anomalous term in the Galilean algebra is
obtained which can be eliminated by demanding conditions on the
Green function. Finally, the axial gauge is also treated by
Dirac's method. 
 Galilean symmetry is preserved in this method. Comparisions
with the  symplectic approach reveal some interesting features.
\end{abstract}

\newpage
\section{Introduction}
Systems of point particles carrying non-abelian charge
interacting with a non-abelian gauge potential have been
considered over the last two decades [1]. Similar models in 2+1
dimensions, where the kinetic term of the gauge field is given by
the Chern-Simons three form instead of the usual Yang-Mills
piece, have been actively investigated in recent years [2-5]. In
this context it is interesting to note that it is possible to
construct models which are Galilean invariant [6-10] rather than
Poincare invariant. This is because the Chern Simons term does
not have an elementary photon associated with it so that the
Bargmann super-selection rule can be accomodated. Purely
Galilean-invariant models are useful to study problems which are
difficult when analysed within the full formalism of special
relativity.

An important issue in the context of theories involving
non-abelian Chern-Simons term is the study of relevant space-time
symmetries associated with either Galilean or Poincare
transformations. For instance it was claimed [2] that (classical)
Poincare covariance gets violated in a theory where the
non-abelian Chern-Simons term is coupled to fermions. The
calculations were done in the axial gauge which enabled the
elimination of the gauge degrees of freedom in favour of the
matter variables. Alternatively, it has been shown by one of us
[5] that by formulating the model in Dirac's [11] constrained
approach which retains all degrees of freedom, the (classical)
Poincare covariance is preserved. It is thus clear that the
issue of symmetries is rather subtle and requires a thorough and
systematic investigation. Indeed since Chern-Simons matter
systems are constrained systems, it is possible to discuss
different formalisms depending on how one accounts for the
constraints. If the results obtained by these distinct formalisms
agree, definitive conclusions emerge. Alternatively, if there is
disagreement it leads to new findings or fresh insights into the
models under investigation. We shall reveal these features
explicitly for a non-relativistic model involving the non-abelian
Chern-Simons term [4].

In section II we take the nonrelativistic model of Bak,Jackiw and Pi [4] and
consider its gauge independent formulation [12-14]. The various
space time generators are defined. The closure of the Galilean
algebra on the constraint surface is then demonstrated. Sections
III and IV comprise the reduced space formalism whereby the gauge
freedom associated with the first class constraints is
eliminated. In section III this is done in the symplectic
approach [4, 15] by choosing the axial gauge. The gauge degrees
of freedom are eliminated in favour of the matter degrees by
solving the Gauss constraint. The space time generators are
defined in the axial gauge. Interestingly, it is found that the
complete Galilean algebra does not close in general. Specifically,
the bracket of the angular momentum with the hamiltonian does not
vanish ; rather it is proportional to a boundary term. This term
vanishes provided some additional restrictions on the Green functions
are imposed. 
 The reduced space discussed in section IV, on the
contrary, is obtained by following Dirac's [11] constrained
analysis. The closure of the full Galilean algebra is
demonstrated without any condition, as was necessary in the
previous (symplectic) approach. We then discuss the origin of
this difference between the two gauge fixed approaches. It is
shown to be related to a boundary term involving the square of
the connection. This boundary term, incidentally, occurred earlier
in the anomalous bracket involving the angular momentum and the
hamiltonian in the symplectic approach. 
Section V contains some concluding remarks.

\section{Gauge independent formulation of the model}
The Lagrangian comprises the Schrodinger Lagrangian minimally
coupled with the non abelian Chern-Simons term [4],
\begin{equation}
{\cal L} = i\psi^{+}D_{\circ}\psi -
\frac{1}{2}(D_{i}\psi)^{+}(D_{i}\psi) -
k\epsilon_{\alpha\beta\gamma}tr(A^{\alpha}\partial^{\beta}A^{\gamma}+\frac{2}{3}A^{\alpha}A^{\beta}A^{\gamma})
\end{equation}
where the covariant derivative is defined by,
\begin{equation}
D_{\mu} = \partial_{\mu}+A_{\mu}
\end{equation}
and $A_{\mu} = A_{\mu a} T^{a}$ with the antihermitian matrices
$T^{a}$ normalised as,
\begin{equation}
tr(T^{a}T^{b}) = - \frac{1}{2}g^{ab}
\end{equation}
where $g^{ab}$ is the metric [4] in group space.
The schrodinger field $\psi$ is an N-component column vector in a
certain representation of $T^{a}$. The Lagrangian (1) can be
written in a canonical form by working out the traces,
\begin{equation}
{\cal L} = i\psi^{+}\dot{\psi} -
\frac{k}{2}\epsilon_{ij}A_{i}^{a}\dot{A}_{j}^{a} -
\frac{1}{2}(\partial_{i}\psi^{+}-\psi^{+}A_{i})(\partial_{i}\psi+A_{i}\psi)+A_{\circ
a}G^{a}
\end{equation}
where
\begin{equation}
G^{a} =
i\psi^{+}T^{a}\psi+\frac{k}{2}\epsilon_{ij}(2\partial_{i}A^{a}_{j}+f^{abc}A^{b}_{i}A^{c}_{j})
\end{equation}
Since ${\cal L}$ has now been expressed in the desired canonical form, it
is simple to read-off the relevant brackets\footnote{All brackets
are referred to at equal times} using the symplectic  approach
[4, 15],
\begin{equation}
\{\psi_{n}(\vec{x}), \psi^{*}_{m}(\vec{x'})\} =
-i\delta_{nm}\delta(\vec{x}-\vec{x'})
\end{equation}
\begin{equation}
\{A_{i}^{a}(\vec{x}), A_{j}^{b}(\vec{x'}\} =
\frac{1}{k}\epsilon_{ij}g^{ab}\delta(\vec{x}-\vec{x'})
\end{equation}
\begin{equation}
\{A_{\circ}^{a}(\vec{x}), \pi_{\circ}^{b}(\vec{x'})\} =
g^{ab}\delta(\vec{x}-\vec{x'})
\end{equation}
where $\pi_{\circ}$ is the momentum conjugate to $A_{\circ}$. All
other brackets vanish. It is clear from (4) that $A^{a}_{\circ}$
is a Lagrange multiplier which enforces the constraint,
\begin{equation}
G^{a} \approx 0
\end{equation}
This constraint is just the analogue of the usual Gauss
constraint in electrodynamics, being the generator of the time
independent non-abelian gauge transformations. Using the brackets
(6-8), it is straightforward to verify this property,
\begin{equation}
\int d^{2}\vec{x} \alpha_{a}(\vec{x}) \{\psi(\vec{x'}),
G^{a}(\vec{x})\} = \alpha(\vec{x'})\psi(\vec{x'})
\end{equation}
\begin{equation}
\int d^{2}\vec{x} \alpha_{b}(\vec{x}) \{A^{a}_{i}(\vec{x'}),
G^{b}(\vec{x})\} =
-\partial_{i}\alpha^{a}(\vec{x'})+f^{abc}\alpha_{b}(\vec{x'})A_{ic}(\vec{x'})
\end{equation}
According to Dirac's [11] classification, therefore, $G^{a}(x)$
is a first class constraint. Indeed it is easy to obtain the
involutive algebra,
\begin{equation}
\{G^{a}(\vec{x}), G^{b}(\vec{x'})\} =
f^{ab}_{c}G^{c}(\vec{x})\delta (\vec{x}-\vec{x'})
\end{equation}

The equations of motion obtained from (1) are found to be,
\begin{equation}
iD_{\circ}\psi = - \frac{1}{2} D_{i} D_{i} \psi
\end{equation}
\begin{equation}
\frac{k}{2}\epsilon_{\alpha\beta\gamma}F^{\beta\gamma} =
J_{\alpha}
\end{equation}
where,
\begin{equation}
J_{\circ} = T_{a}J_{\circ}^{a} = T_{a}(-i\psi^{+}T^{a}\psi)
\end{equation}
\begin{equation}
J_{i} = T_{a}J_{i}^{a} =
-\frac{1}{2}T_{a}[\psi^{+}T^{a}D_{i}\psi-(D_{i}\psi)^{+}T^{a}\psi]
\end{equation}
are the non-abelian charge density and spatial current density
respectively. As usual, the time-component of (14) yields the
Gauss constraint (9).

Going over to the hamiltonian formalism we observe that the
momenta canonically conjugate to the Lagrange multiplier
$A_{\circ}$ is a constraint,
\begin{equation}
\pi^{a}_{\circ} \approx 0
\end{equation}
This, together with $G^{a}$,form the complete set of constraints.
The relations,
\begin{equation}
\{\pi_{\circ}^{a}(\vec{x}), G^{b}(\vec{x'})\} =
\{\pi^{a}_{\circ}(\vec{x}),\pi_{\circ}^{b}(\vec{x'})\} = 0
\end{equation}
along with (12) constitute the full involutive algebra among the
constraints. The canonical hamiltonian is immediately written on
inspecting (4),
\begin{equation}
H_{c} = \int
d^{2}\vec{x}(\frac{1}{2}(D_{i}\psi)^{+}(D_{i}\psi)-A_{\circ}^{a}G_{a})
\end{equation}
Using (6-8) it is easy to verify that $H_{c}$ correctly generates
the equations of motion,
\begin{equation}
\partial_{\circ}\psi_{n} = \{\psi_{n}, H_{c}\}
\end{equation}
\begin{equation}
\partial_{\circ}A^{a}_{i} = \{A^{a}_{i}, H_{c}\}
\end{equation}

Let us next discuss the symmetries under various space-time
transformations. Consider an infinitesimal transformation,
\begin{equation}
x_{\mu} \rightarrow x'_{\mu} = x_{\mu} + \delta x_{\mu}
\end{equation}
\begin{equation}
\phi_{n}(x) \rightarrow \phi'_{n}(x') = \phi_{n}(x) +
\delta\phi_{n}(x)
\end{equation}
with,
\begin{equation}
\delta x_{\mu} = \wedge_{\mu\nu}\delta\omega^{\nu}
\end{equation}
\begin{equation}
\delta \phi_{n} = \Phi_{n\nu}\delta\omega^{\nu}
\end{equation}
where $\phi_{n}(x)$ generically denotes the fields in the
Lagrangian and $\nu$ can be a single or double index. Then the
invariance of any Lagrangian under the above transformations
leads to a conserved current [16],
\begin{equation}
J_{\mu\nu} =
\frac{\partial{\cal L}}{\partial(\partial^{\mu}\phi_{n})}\Phi_{n\nu}
- \theta_{\mu\sigma}\wedge^{\sigma}_{\nu}
\end{equation}
where,
\begin{equation}
\theta_{\mu\sigma} = \frac{\partial
{\cal L}}{\partial(\partial^{\mu}\phi_{n})} \partial_{\sigma}\phi_{n} -
{\cal L} g_{\mu\sigma}
\end{equation}

With this general input it is straightforward to obtain the
various Galileo generators of the present model. For example,
under space translations,
\begin{equation}
x_{i} \rightarrow x'_{i} = x_{i} - \delta\omega_{i}
\end{equation}
\begin{equation}
x_{\circ} \rightarrow x'_{\circ} = x_{\circ}
\end{equation}
the fields do not transform $(\Phi_{n\nu} = 0)$ and the relevant
generator is given by
\begin{eqnarray}
P_{i} &= &\int d^{2}\vec{x} \theta_{\circ i}(\vec{x})\nonumber \\
      &= &\int d^{2}\vec{x}(i\psi^{+}\partial_{i}\psi -
      \frac{k}{2}\epsilon_{kj}A^{a}_{k}\partial_{i}A_{ja})
\end{eqnarray}
Once again using (6-8) the normal transformation properties for
the fields may be checked,
\begin{equation}
\{A_{k}(\vec{x}), P_{i}\} = \partial_{i}A_{k}(\vec{x})
\end{equation}
\begin{equation}
\{\psi(\vec{x}), P_{i}\} = \partial_{i}\psi(\vec{x})
\end{equation}
Similarly, under infinitesimal spatial rotations with angle
$\theta$,
\begin{equation}
t' = t
\end{equation}
\begin{equation}
x'_{i} = x_{i} + \theta\epsilon_{ij}x_{j}
\end{equation}
the fields transform as,
\begin{equation}
A'_{\circ}(x') = A_{\circ}(x), \psi'(x') = \psi(x)
\end{equation}
\begin{equation}
A'_{i}(x') = A_{i}(x)+\theta\epsilon_{ij}A_{j}(x)
\end{equation}
Comparing with (24,25) we find,
\begin{equation}
\delta x_{i} = \theta\epsilon_{ij}x_{j}, \wedge_{ijk} =
\delta_{ij}x_{k}, A^{a}_{ijk} = \delta_{ij}A^{a}_{k}
\end{equation}
The rotation generator after an antisymmetrisation now follows
from (26),
\begin{eqnarray}
J_{ij} &= &\int d^{2}\vec{x} J_{\circ[ij]} \nonumber \\
       &= &\int d^{2}\vec{x}(x_{[i}\theta_{\circ
       j]}+\frac{k}{2}\epsilon_{[im}A_{ma}A^{a}_{j]})
\end{eqnarray}
Since there is only one component, we may express this as,
\begin{equation}
J = \int d^{2}\vec{x}(\epsilon_{ij}x_{i}\theta_{\circ j} +
\frac{k}{2}A_{ja}A^{a}_{j})
\end{equation}
The basic fields obey covariant transformation laws,
\begin{equation}
\{\psi(\vec{x}), J\} =
\epsilon_{ij}x_{i}\partial_{j}\psi(\vec{x})
\end{equation}
\begin{equation}
\{A^{a}_{i}(\vec{x}), J\} =
\epsilon_{jk}x_{j}\partial_{k}A^{a}_{i}(\vec{x})+\epsilon_{ij}A^{a}_{j}(\vec{x})
\end{equation}
Finally, we come to the Galileo boosts,
\begin{equation}
x_{i} \rightarrow x'_{i} = x_{i} - \vartheta_{i}t
\end{equation}
The fields transform as,
\begin{equation}
\psi'(x',t') = \psi(x,t) - i\vartheta_{i}x_{i}\psi(x,t)
\end{equation}
\begin{equation}
A_i^\prime(x',t')= A_i(x,t)
\end{equation}
\begin{equation}
A'_{\circ}(x',t') = A_{\circ}(x,t) + \vartheta_{i}A_{i}(x,t)
\end{equation}
It can be verified that the action corresponding to (1) is
invariant under these transformations. Comparing (42-45) with
(24,25) yields the correspondence,
\begin{equation}
\wedge_{ij} = -t\delta_{ij}, \Phi_{nj} = -i\psi_{n}x_{j}
\end{equation}
so that the boost generator may be obtained from (26,27) as,
\begin{eqnarray}
K_{i} &= &\int d^{2}\vec{x} J_{\circ i} \nonumber \\
      &= &\int(\frac{\partial L}{\partial \dot{\psi}_{n}} \Phi_{ni} -
      \theta_{\circ j} \wedge_{ji})d^{2}\vec{x} \nonumber \\
      &= &t P_{i} + \int d^{2}\vec{x} x_{i} \psi^{+}\psi
\end{eqnarray}
Under these boosts the basic fields have the usual transformation
properties,
\begin{equation}
\{\psi(\vec{x}), K_{i}\} = t\partial_{i}\psi(\vec{x}) -
ix_{i}\psi(\vec{x})
\end{equation}
\begin{equation}
\{A_{j}(\vec{x}), K_{i}\} = t\partial_{i}A_{j}(\vec{x})
\end{equation}
We have thus shown that the basic fields transform covariantly
under all the (Galilean) space-time generators. Consequently it is
expected that the complete Galilean algebra ought to be satisfied.
Indeed an explicit computation reveals that,
\begin{equation}
\{P_{i}, P_{j}\} = \{P_{i}, H_{c}\} = \{K_{i}, K_{j}\} = 0
\end{equation}
\begin{equation}
\{P_{i}, J\} = \epsilon_{ij} P_{j}
\end{equation}
\begin{equation}
\{K_{i}, J\} = \epsilon_{ij} K_{j}
\end{equation}
\begin{equation}
\{P_{i}, K_{j}\} = \delta_{ij} \int d^{2}\vec{x}\psi^{+}\psi =
\delta_{ij}M
\end{equation}
\begin{equation}
\{H_{c}, K_{i}\} = P_{i} + \int d^{2}\vec{x} A^{a}_{i} G_{a} \approx
P_{i}
\end{equation}
\begin{equation}
\{J, H_{c}\} = \epsilon_{ij} \int d^{2}\vec{x} x_{i}
A^{a}_{\circ}\partial_{j}G_{a} \approx 0
\end{equation}
The last two brackets reduce to the conventional result on the
constraint surface. Thus, on this surface classical Galilean
covariance of the model has been demonstrated. An identical
conclusion also holds in the abelian model [10].

The last part of this section is devoted to show that the
generators entering in the above (Galilean) algebra are all gauge
invariant. In that case these generators can be regarded as
physical entities. Using the basic brackets (6-8), the algebra of
the Gauss constraint (5) with the various generators may be
explicitly calculated to yield,
\begin{equation}
\{P_{i}, G^{a}(\vec{x})\} = -\partial_{i}G^{a}(\vec{x}) \approx 0
\end{equation}
\begin{equation}
\{H_{c}, G^{a}(\vec{x})\} = 0
\end{equation}
\begin{equation}
\{J, G^{a}(\vec{x})\} =
-\epsilon_{ij}x_{i}\partial_{j}G^{a}(\vec{x}) \approx 0
\end{equation}
\begin{equation}
\{K_{i}, G^{a}(\vec{x})\} = -t\partial_{i}G^{a}(\vec{x}) \approx 0
\end{equation}
Thus all the generators are found to be gauge invariant on the
constraint surface defined by (9). This completes the gauge
independent formulation of the model. The independent canonical
pairs are $(A_{1},A_{2})$ and $(\psi, \psi^{+})$. Classical
Galilean algebra is satisfied. Furthermore gauge invariance of the
relevant generators implies that this algebra should also be
preserved in a gauge fixed analysis. Nevertheless it is
interesting and instructive to explicitly perform the gauge fixed
computations. This will provide fresh insights into the model.

\section{Gauge fixed formulation : The symplectic approach}
The basic idea of this formulation is to work in a reduced space by
eliminating the gauge freedom. There are different ways to achieve
this target. In the symplectic approach [15] one explicitly solves
the Gauss constraint (9) by imposing an additional (gauge)
condition, thereby eliminating the gauge degrees of freedom in
favour of the matter variables. But whatever gauge condition is
chosen, the resulting solution is nonlocal since it involves the
inversion of a derivative. In this sense the concept of a local
action breaks down. A particularly effective gauge choice is the
axial gauge,
\begin{equation}
A_{1}^{a} = 0
\end{equation}
since it linearises the Gauss constraint,
\begin{equation}
G^{a} = k\partial_{1}A^{a}_{2} - J^{a}_{\circ} = 0
\end{equation}
so that the other component of the gauge field is given by,
\begin{equation}
A^{a}_{2} = \frac{1}{k}\int
d^{2}\vec{x'}G(\vec{x}-\vec{x'})J^{a}_{\circ}(\vec{x'})
\end{equation}
where $G(\vec{x}-\vec{x'})$ is the Green function,
\begin{equation}
\partial_{1}G(\vec{x}-\vec{x'}) = \delta(\vec{x}-\vec{x'})
\end{equation}
The algebra of the gauge sector is now completely governed by the
basic bracket (6) in the matter sector. Using (60) and (62) it
follows,
\begin{equation}
\{A^{a}_{1}(\vec{x}), A_{1}^{b}(\vec{x'})\} = \{A_{1}^{a}(\vec{x}),
A_{2}^{b}(\vec{x'})\} = 0
\end{equation}
\begin{equation}
\{A_{2}^{a}(\vec{x}), A_{2}^{b}(\vec{x'})\} = -\frac{1}{k^{2}}\int
d^{2}\vec{y}G(\vec{x}-\vec{y})G(\vec{x'}-\vec{y})f^{abc}J_{\circ
c}(\vec{y})
\end{equation}
Likewise it is easy to obtain the algebra of the  mixed sector,
\begin{equation}
\{A^{a}_{1}(\vec{x}), \psi_{n}(\vec{x'})\} = 0
\end{equation}
\begin{equation}
\{A^{a}_{2}(\vec{x}), \psi_{n}(\vec{x'})\} =
\frac{1}{k}(T^{a}\psi(\vec{x'}))_{n}G(\vec{x}-\vec{x'})
\end{equation}
The algebra involving $A^{a}_{\circ}$ is inconsequential since it is
a Lagrange multiplier and not a dynamical variable. Note that there
is an important subtlety in the solution (62). It does not
represent a unique solution for $A^{a}_{2}$. There is an
arbirtrariness because if $A^{a}_{2}(x)$ is a solution then
$A'^{a}_{2}(x) = A_{2}^{a}(x) + f^{a}(x_{\circ},x_{2})$ is also a
solution. On the other hand there is a residual gauge freedom that
survives the axial gauge (60) [5],
\begin{equation}
A^{a}_{\mu}(x)\rightarrow A'^{a}_{\mu}(x) =
A^{a}_{\mu}(x)+\partial_{\mu}\alpha^{a}(x_{\circ},x_{2})+f^{abc}A_{\mu
b}(x)\alpha_{c}(x_{\circ},x_{2})
\end{equation}
In the abelian theory it is possible, by choosing $f^{a} =
\partial_{2}\alpha^{a}$, to account for the residual gauge freedom
and regard (62) as a unique solution for the gauge field. For the
nonabelian theory at hand,  however, the presence of the extra
piece in (68) complicates matters. Indeed if we take,
\begin{equation}
f^{a}(x_{\circ},x_{2}) =
\partial_{2}\alpha^{a}(x_{\circ},
x_{2})+f^{abc}A_{2b}(x)\alpha_{c}(x_{\circ}, x_{2})
\end{equation}
we find that while the l.h.s. depends on $(x_{\circ}, x_{2})$ only,
the r.h.s. depends on all $x$, so that it becomes impossible to
find a solution for $f^{a}$. The arbitrariness in (62), therefore,
persists just as the residual gauge freedom due to (68) remains.

The implementation of a specific gauge is known to modify the
manifest covariant transformation of the basic fields [17]. For
instance in the radiation gauge the boost law is found to be
altered [17]. In the axial gauge, on the other hand, manifest
rotational symmetry is violated. This implies that the
transformation (41) under rotations will be modified to preserve
the gauge condition (60). Since manifest covariance is spoilt it
becomes imperative to verify the Galilean symmetry by working out
the algebra (50-55) involving the gauge invariant generators. A
detailed analysis shows that apart from one exception the complete
Galilean algebra (50-55) is reproduced. The only nontrivial bracket
is given by,
\begin{equation}
\{J, H_{c}\} = \frac{1}{k}\int d^{2}\vec{x}
d^{2}\vec{y}d^{2}\vec{z}\{G(\vec{x}-\vec{y})G(\vec{y}-\vec{z})-G(\vec{x}-\vec{z})G(\vec{y}-\vec{z})\}f_{abc}A^{a}_{2}(\vec{x})J_{2}^{b}(\vec{y})J_{\circ}^{c}(\vec{z})
\end{equation}
where the current $J_{2}^{b}$ and charge density $J_{\circ}^{c}$
are defined in (15-16), and $A^{a}_{2}$ is given in (62). It is
possible to simplify the r.h.s. of (70) by replacing
$J_{\circ}^{c}$ using (61),
\begin{equation}
\{J, H_{c}\} = \int
d^{2}\vec{x}d^{2}\vec{y}d^{2}\vec{z}\{G(\vec{x}-\vec{y})G(\vec{y}-\vec{z})-G(\vec{x}-\vec{z})G(\vec{y}-\vec{z})\}f_{abc}A_{2}^{a}(\vec{x})J_{2}^{b}(\vec{y})\partial_{1}A_{2}^{c}(\vec{z})
\end{equation}
Using (63), one can further simplify to obtain,
\begin{equation}
\{J, H_{c}\} = -\int
d^{2}\vec{x}d^{2}\vec{y}d^{2}\vec{z}\partial^{z}_{1}\{G(\vec{x}-\vec{z})G(\vec{y}-\vec{z})A^{c}_{2}(\vec{z})\}f_{abc}A^{a}_{2}(\vec{x})J_{2}^{b}(\vec{y})
\end{equation}

As pointed out in the previous section the algebra of the gauge
invariant generators must be independent of the choice of gauge.
Since the complete Galilean algebra was demonstrated earlier, it
implies that $\{J, H_{c}\}$ must vanish in the axial gauge. The
r.h.s. of (72) shows that this is not true in general. A simple
way to establish compatibility is to demand that the boundary
term vanishes, i.e.,
\begin{equation}
\int d^{2}\vec{z}
\partial^{z}_{1}\{G(\vec{x}-\vec{z})G(\vec{y}-\vec{z})A^{c}_{2}(\vec{z})\}
= 0
\end{equation}
The above relation gives a restriction on the connection
$G(\vec{x}-\vec{y})$. Note that this connection appears squared
which must be regularised [4, 18] to make it meaningful. The
regularisation must be such that the above condition (73) is
satisfied. In that case the complete Galilean algebra is
reproduced. It is useful to compare this analysis with Dirac's
gauge fixed approach which is given below.

\section{Gauge fixed formulation : Dirac's approach}
In contrast to the symplectic approach the hamiltonian analysis
of Dirac [11] distinguishes between first class and second class
constraints. The gauge freedom generated by the first class
constraint $G^{a} \approx 0$ is eliminated by initially choosing
a gauge $\chi^{b} \approx 0$ so that,
\begin{equation}
\det \vert\vert \{G^{a}, \chi^{b}\}\vert\vert \not= 0
\end{equation}
Then the complete set of constraints $G^{a} \approx 0, \chi^{b}
\approx 0$ becomes second class which can be strongly implemented
by working with Dirac (star) brackets,
\begin{equation}
\{\phi^{a}(\vec{x}), \phi^{b}(\vec{y})\}^{*} =
\{\phi^{a}(\vec{x}), \phi^{b}(\vec{y})\} - \int
d^{2}zd^{2}z'\{\phi^{a}(\vec{x}),\Omega^{c}(\vec{z})\}
\Omega^{-1}_{cd}(\vec{z},\vec{z'})\{\Omega^{d}(\vec{z'}),\phi^{b}(\vec{y})\}
\end{equation}
where $\Omega^{-1}_{cd}$ is the inverse of the matrix defined by
the Poisson brackets $\{\Omega_{c}, \Omega_{d}\}$ of the complete
set of constraints $\Omega_{c} = G_{c}, \chi_{c} \approx 0$. The
ordinary brackets in (74) merely refer to the fundamental
brackets (6-8).

It is worthwhile to highlight some of the fundamental
distinctions between the implementation of constraints in the
symplectic [4, 15] and Dirac [11] approaches. Contrary to the
symplectic case, all the degrees of freedom (either gauge or
matter) are retained in the Dirac analysis. There is no need for
an explicit solution of the Gauss constraint (61) leading to
the non-local structure (62). This also avoids the inherent
arbitrariness in the solution (62, 63).

The next step is to compute the Dirac brackets among the basic
fields in the axial gauge. The matrix of the Poisson brackets of
the constraints is given by,
\begin{equation}
\Omega^{ab}_{ij}(\vec{x},\vec{y}) =
\vert\vert\{\Omega^{a}_{i}(\vec{x}),
\Omega^{b}_{j}(\vec{y})\}\vert\vert = - \left(\begin{array}{cc} 0
&\partial^{x}_{1} \\ \partial^{x}_{1} &0 \end{array}\right)
\delta(\vec{x}-\vec{y})g^{ab}
\end{equation}
where $\Omega^{a}_{1} = A^{a}_{1} \approx 0$ and $\Omega^{a}_{2}
= G^{a} \approx 0$. The corresponding inverse matrix is found to
be,
\begin{equation}
(\Omega^{ab}_{ij}(x,y))^{-1} = \left(\begin{array}{cc} 0 &-1 \\ -1
&0\end{array}\right) G(\vec{x}-\vec{y})g^{ab}
\end{equation}
where the connection has been defined in (63). From the basic
brackets (6-8) and using the definition of Dirac brackets in (75)
we find the gauge fixed algebra,
\begin{equation}
\{\psi_{n}(\vec{x}), A^{a}_{2}(\vec{y})\}^{*} = \frac{1}{k}
G(\vec{x} - \vec{y})[T^{a}\psi(\vec{x})]_{n}
\end{equation}
\begin{equation}
\{A^{a}_{2}(\vec{x}), A^{b}_{2}(\vec{y})\}^{*} = \frac{1}{k}
f^{abc} G(\vec{x} - \vec{y}) (A_{2c}(\vec{x}) - A_{2c}(\vec{y}))
\end{equation}
The brackets with $A_{1}^{a}(\vec{x})$ vanish as expected from
the gauge condition. Note that the second relation preserves the
antisymmetry of the bracket under the simultaneous exchange $x
\leftrightarrow y, a \leftrightarrow b$.

Let us now compare the Dirac algebra with the corresponding
symplectic algebra. Although the bracket (78) agrees with (67),
the bracket (79) has a different structure from (65). Thus, at
the level of the basic algebra, we find a distinction between the
two approaches. It now takes only a little effort to show that
the difference between (65) and (79) is just the boundary term in
the l.h.s. of (73). Using (61), the bracket (65) reduces to the
following,
\begin{eqnarray}
\{A^{a}_{2}(\vec{x}), A_{2}^{b}(\vec{x'})\} &= &-\frac{1}{k}\int
d^{2}\vec{y}G(\vec{x}-\vec{y})G(\vec{x'}-\vec{y})f^{abc}\partial_{1}A_{2c}(\vec{y})
\nonumber \\
        &=&\frac{f^{abc}}{k}\int
        d^{2}\vec{y}\partial^{y}_{1}[G(\vec{x}-\vec{y})G(\vec{x'}-\vec{y})A_{2c}
(\vec{y})]\nonumber\\
&+&\frac{1}{k}f^{abc}G(\vec{x'}-\vec{x})(A_{2c}(\vec{x})-A_{2c}(\vec{x'}))
        \nonumber \\
        &= &\{A^{a}_{2}(\vec{x}),
        A_{2}^{b}(\vec{x'})\}^{*}+\frac{f^{abc}}{k}\int
        d^{2}\vec{y}\partial^{y}_{1}[G(\vec{x}-\vec{y})G(\vec{x'}-\vec{y})A_{2c}(\vec{y})]
\end{eqnarray}
where, in going from the first to the second line, we have used
(63). Thus, as announced, the difference between the symplectic
and Dirac algebras is proportional to the boundary term in (73).
If we impose the condition (73) then the two results agree.
Finally, using the Dirac brackets (78, 79), it is possible to
establish the validity of the complete Galilean algebra (without
any restrictions) including the bracket $\{J, H_{c}\}^{*}$ which
previously yielded an anomalous structure (72) in the symplectic
approach. This is not surprising because the anomalous structure
in (72) is precisely compensated by the difference in the basic
bracket $\{A^{a}_{2}, A^{b}_{2}\}$ (80) in the two approaches.

\section{Conclusions}
We have investigated by different approaches the (classical)
Galilean symmetry in a nonrelativistic model involving the
coupling of nonabelian Chern-Simons term to matter fields [4].
Since the model is a constrianed system there are different
formulations depending on how one accounts for the constraints. A
conceptually clean and elegant way of doing this is to work in
the gauge independent formalism [12-14]. The various Galilean
generators are defined. It is also verified that on the
constraint surface, these generators are gauge invariant. The
basic fields are found to transform covariantly under the different
space-time generators. The Galilean algebra is reproduced on the
constraint surface. Since this algebra involves (physical) gauge
invariant quantities, it implies that the algebra should be
preserved in any gauge fixed compuatation. Two distinct
approaches to gauge fixing have been considered in this paper.
The first is the symplectic approach [15] whereby the Gauss
constraint is explicitly solved in the axial gauge. The gauge
degrees of freedom are eliminated in favour of the matter
variables. Since the process involves the inversion of a
derivative, the solution for the gauge field is nonlocal. It is
found that, except for $\{J, H_{c}\}$, the Galilean algebra is
preserved. The bracket $\{J, H_{c}\}$ is anomalous; it is in fact
proportional to a boundary term involving the square of the
connection. To establish compatibility with the Galilean
covariance the boundary term ought to vanish which, therefore,
imposes restrictions on the connection. The gauge fixed analysis
is next repeated in the Dirac [11] formalism. Contrary to the
symplectic approach explicit solution of the Gauss constraint is
not necessary so that nonlocal expressions do not occur. The
complete Galilean algebra (including the result for $\{J,
H_{c}\}$) is valid, without any restrictions on the connection.
The difference in the result obtained in the Dirac and symplectic
approaches is attributed to the fact that the basic bracket
$\{A^{a}_{2}(\vec{x}), A_{2}^{b}(\vec{x'})\}$ is different in the
two approaches. This difference is just proportional to the
previously mentioned boundary term occurring in $\{ J,H_c\}$
computed in the symplectic approach.
Not surprisingly, therefore,
the Dirac approach leads to a straightforward validity of the
Galilean algebra. It may be worthwhile to consider the effects of
ordering so that an analysis of the (quantum) Galilean covariance
can be pursued. The gauge independent formulation should be the
likely starting point since, contrary to gauge fixed approaches,
the algebra in the mixed sector is trivial. Moreover,
identification of the independent canonical pairs $(A_{1}^{a},
A_{2}^{a}), (\psi, \psi^{*})$ is also clean. These and other
related issues are currently under investigation.

\end{document}